\documentclass[a4paper]{article}

\usepackage[english]{babel}
\usepackage[utf8x]{inputenc}
\usepackage{amsmath}
\usepackage{amsfonts}
\usepackage{amsthm}
\usepackage{graphicx}
\usepackage{color}
\usepackage{authblk} 
\usepackage[margin=3cm]{geometry} 
\usepackage{bm} 
\usepackage{appendix}

\title{Effect of diluted connectivities on cluster synchronization of adaptively coupled oscillator networks}

\author[1]{Simon Vock}
\author[1,2]{Rico Berner}
\author[2]{Serhiy Yanchuk}
\author[1,3,4]{Eckehard Sch\"oll}
\affil[1]{Institute of Theoretical Physics, Technische Universit\"at Berlin, Hardenbergstra\ss e~36, 10623 Berlin, Germany}
\affil[2]{Institute of Mathematics, Technische Universit\"at Berlin, Stra\ss e des 17. Juni 136, 10623 Berlin, Germany}
\affil[3]{Bernstein Center  for  Computational  Neuroscience  Berlin, Humboldt-Universit\"at, Philippstra\ss e  13,  10115  Berlin,  Germany}
\affil[4]{Potsdam Institute for Climate Impact Research, Telegrafenberg A 31, 14473 Potsdam, Germany}
\date{} 

\begin{document}
\maketitle
\begin{abstract}
Synchronization in networks of oscillatory units is an emergent phenomenon present in various systems, such as biological, technological, and social systems. Many real-world systems have adaptive properties, meaning that their connectivities change with time, depending on the dynamical state of the system. Networks of adaptively coupled oscillators show various synchronization phenomena, such as hierarchical multifrequency clusters, traveling waves, or chimera states. While these self-organized patterns have been previously studied on all-to-all coupled networks, this work extends the investigations towards more complex networks, analyzing the influence of random network topologies for various degrees of dilution of the connectivities. Using numerical and analytical approaches, we investigate the robustness of multicluster states on networks of adaptively coupled Kuramoto-Sakaguchi oscillators against the random dilution of the underlying network topology. Further, we utilize the master stability approach for adaptive networks in order to highlight the interplay between adaptivity and topology. 
\end{abstract}

\section{Introduction}\label{sec:intro}

In general terms, a complex dynamical network is a set of dynamical units (nodes) with connections between them (links), representing a relation or interaction among the individual elements. In nature as well as in technology, complex dynamical networks provide a framework with a broad range of applications in physics, chemistry, biology, neuroscience, economy, social science, and many more~\cite{NEW03,BOC18}. 

Collective behavior in dynamical networks is the emergent phenomenon of spontaneous ordered dynamics. One example of particular importance is synchronization~\cite{PEC97,PIK01,STR01a,YAN03b,ARE08,BOC18}. First recognized by Huygens in the $17$th century~\cite{HUY73}, synchronization phenomena of coupled oscillators are of great interest in science, nature, engineering, and social life. Depending on the dynamical properties of a system, diverse synchronization patterns of varying complexity have been observed, such as complete synchronization~\cite{KUR84}, cluster synchronization \cite{YAN01a,SOR07,DAH12,MIY13,ZHA20}, and various forms of partial synchronization~\cite{KUR02a,ABR04,MOT10,PAN15,SCH16b,SAW18c,OME19c,AND20,DRA20,SCH20b,ZAK20,SCH20}. Synchronization patterns are believed to play an important role in neural networks, for instance in the context of cognition and learning \cite{SIN99,WAN10e,FEL11} as well as in pathological conditions such as Parkinsons's disease \cite{HAM07,SCH10h,KRO20,KRO20a} or epilepsy~\cite{LEH09,MOR00,JIR13,JIR14,ROT14,AND16,CHO18,GER20}.

A central question in the study of synchronization on complex networks is whether such behavior is stable. A powerful analytic framework to study the stability of synchronized states is the master stability approach~\cite{PEC98}. Since its introduction, this approach has been extended to various types of networks, such as multilayer networks~\cite{BRE18,BER20}, networks with time-delays \cite{CHO09,FLU10b,GRE10,LEH11, DAH12, FLU12, KEA12, WIL14, KYR14, LEH15b, HUD16, BOE20}, hypernetworks \cite{SOR12a,MUL20}, and very recently to adaptive networks~\cite{BER20b}. The master stability approach allows for the separation of the effects of the local node dynamics from the effects of the network topology. This can be used to draw general conclusions on the stability of dynamical systems by analyzing the eigenvalues of the network connectivity matrix.

Most of the previous studies have analyzed the dynamical processes that occur on static networks, describing fixed interaction structures that do not change with time. Real-world networks however often change in time, adapting their structure in response to the network state~\cite{GRO08a,AOK09,KAS17}. This type of network is called adaptive or co-evolutionary, and combines topological evolution of the network with dynamics on the network nodes. This behavior appears in many real-world applications. For instance, power grids and traffic networks continuously change to meet the evolving needs of society. Further, adaptive behavior is of great importance in neural networks. These are networks of individual neurons connected by synapses that pass electrical or chemical signals between them. It has been shown that the coupling weights between the individual neurons may be potentiated or depressed, depending on the order of the spike times of post- and presynaptic neuron~\cite{GER96,ABB00,DAN06}. This mechanism called {\em spike timing-dependent plasticity} is believed to play an important role in temporal coding of information in the brain~\cite{CLO10}.

In this article, we investigate the robustness of multicluster states on networks of adaptively coupled Kuramoto-Sakaguchi oscillators against the random dilution of the underlying network topology. By randomly and successively deleting links, we observe a linear dependence of the cluster frequencies on the relative number of deleted links. We explain this linear dependence by a suitable approximation of the oscillators frequencies. We show numerically that the shape of a multicluster state is preserved on networks of different sparsity, ranging from fully coupled to almost uncoupled topologies. Further, we show that the multicluster states observed are multistable, meaning that different multiclusters may emerge for different initial conditions. By utilizing the master stability approach for adaptive networks, we present the effects induced by changes of the network topology on the desynchronization of the phase synchronized state. We show that the resulting desynchronization has strong implications for the emergence of multicluster states.

This article is organized as follows. In section~\ref{sec:model}, we introduce the model used throughout this work. In section~\ref{sec:multicluster}, we discuss the emergence and structural formation of multi-frequency-cluster states. By applying a suitable randomization process on the network connectivity, we investigate the effect of random dilution of links in the connectivity structure on multicluster states in section \ref{sec:dilution}. Further, we apply the master stability approach for adaptive networks to investigate the interplay between nodal dynamics, adaptivity, and a complex connectivity structure in section \ref{sec:msf_phase_osci}. We use these results to describe a desynchronization transition by changes in the network topology in section \ref{sec:desync_top_change}.

\section{Model}\label{sec:model}
	
\begin{figure}
		\centering
		\def\svgwidth{0.8\columnwidth}
\begingroup%
  \makeatletter%
  \providecommand\color[2][]{%
    \errmessage{(Inkscape) Color is used for the text in Inkscape, but the package 'color.sty' is not loaded}%
    \renewcommand\color[2][]{}%
  }%
  \providecommand\transparent[1]{%
    \errmessage{(Inkscape) Transparency is used (non-zero) for the text in Inkscape, but the package 'transparent.sty' is not loaded}%
    \renewcommand\transparent[1]{}%
  }%
  \providecommand\rotatebox[2]{#2}%
  \newcommand*\fsize{\dimexpr\f@size pt\relax}%
  \newcommand*\lineheight[1]{\fontsize{\fsize}{#1\fsize}\selectfont}%
  \ifx\svgwidth\undefined%
    \setlength{\unitlength}{1136.76782803bp}%
    \ifx\svgscale\undefined%
      \relax%
    \else%
      \setlength{\unitlength}{\unitlength * \real{\svgscale}}%
    \fi%
  \else%
    \setlength{\unitlength}{\svgwidth}%
  \fi%
  \global\let\svgwidth\undefined%
  \global\let\svgscale\undefined%
  \makeatother%
  \begin{picture}(1,0.37335648)%
    \lineheight{1}%
    \setlength\tabcolsep{0pt}%
    \put(0,0){\includegraphics[width=\unitlength,page=1]{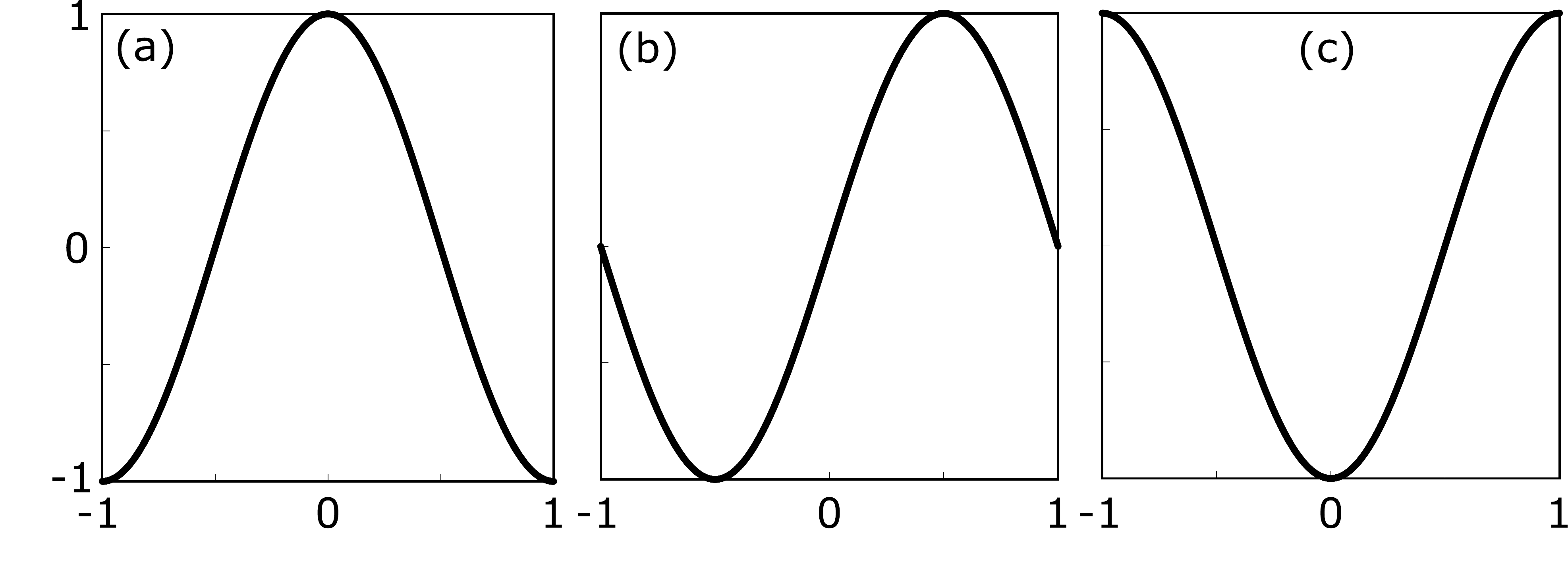}}%
    \put(0.2091315,0.00530903){\makebox(0,0)[t]{\lineheight{1.25}\smash{\begin{tabular}[t]{c}$\Delta\phi/\pi$\end{tabular}}}}%
    \put(0.52899436,0.00530903){\makebox(0,0)[t]{\lineheight{1.25}\smash{\begin{tabular}[t]{c}$\Delta\phi/\pi$\end{tabular}}}}%
    \put(0.84885728,0.00530903){\makebox(0,0)[t]{\lineheight{1.25}\smash{\begin{tabular}[t]{c}$\Delta\phi/\pi$\end{tabular}}}}%
    \put(0.02030834,0.21641222){\rotatebox{90}{\makebox(0,0)[t]{\lineheight{1.25}\smash{\begin{tabular}[t]{c}$-\sin(\Delta\phi+\beta)$\end{tabular}}}}}%
  \end{picture}%
\endgroup%

		\caption{The adaptation function $-\sin(\Delta\phi + \beta)$ used in system \eqref{eq:adapt_phase_osci1}--\eqref{eq:adapt_phase_osci2}, for the parameter values (a) $\beta = -\pi/2$ (Hebbian), (b) $\beta = 0$ (causal, effectively similar to spike timing-dependent plasticity), (c) $\beta = \pi/2$ (anti-Hebbian).
	}
		\label{fig:adaption_function_phaseosci}
	\end{figure}
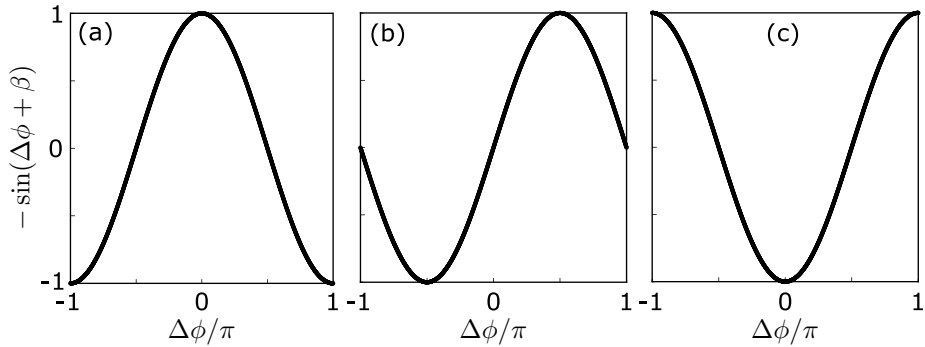

	We consider a network of $N$ Kuramoto-Sakaguchi type phase oscillators with adaptive coupling weights, described by
	\begin{align}
		\dot \phi_i &= \omega - \sigma \sum_{j = 1}^{N} a_{ij} \kappa_{ij} \sin \left( \phi_i - \phi_j + \alpha \right),\label{eq:adapt_phase_osci1} \\
		\dot \kappa_{ij} &= -\epsilon \left( \kappa_{ij} + a_{ij} \sin \left( \phi_i - \phi_j + \beta \right) \right),\label{eq:adapt_phase_osci2}
	\end{align}
where $\phi_i(t)\in[0,2\pi)$ describes the phase of oscillator $i \in \{1,\dots,N\}$, and $\kappa_{ij}(t)\in[-1,1]$ denotes the coupling strength from oscillator $j$ to $i$. The connectivity structure is given by the elements $a_{ij}\in\{0,1\}$ of the adjacency matrix $A$ which is independent of time. Note that self-coupling does not influence the relative dynamics, which is why the $N$ diagonal elements $a_{ii}$ are set to zero in all our simulations. Equation~\eqref{eq:adapt_phase_osci1} describes a network of Kuramoto-Sakaguchi type phase oscillators with a diffusive coupling kernel $\sin\left( \phi_i - \phi_j + \alpha \right)$ scaled by an overall coupling strength $\sigma$, and $\omega$ is the intrinsic frequency. The parameter $\alpha$ can be considered as a phase lag of the interaction between the oscillators~\cite{SAK86} or even related to a synaptic propagation delay~\cite{ASL18a}. Equation~\eqref{eq:adapt_phase_osci2} describes the dynamics of the coupling weights $\kappa_{ij}$, and $\epsilon \ll 1$ is a time-scale separation parameter. We define the adaptation function as $-\sin \left( \phi_i - \phi_j + \beta \right)$, where $\beta$ is a control parameter, enabling us to tune between different plasticity rules that can occur in neural networks, see Fig.~\ref{fig:adaption_function_phaseosci}. For instance, setting $\beta = -\pi/2$ (Fig.~\ref{fig:adaption_function_phaseosci}(a) results in an adaptation function corresponding to a Hebbian-like rule. In this case, the adaptation function is positive and acts towards the increase of the synaptic weights if the phase difference of post- and pre-synaptic neuron are close to each other, i.e., $|\Delta \phi| < \pi/2$, where $\Delta \phi = \phi_i - \phi_j$.
	
	The synchrony of the oscillators at a given time $t$ is typically quantified by the Kuramoto-Daido order parameter \cite{KUR84,DAI92a}. The complex $n$th order parameter for the state $\bm{\phi}(t)= (\phi_1(t),\dots,\phi_N(t))^T$ is defined as
	\begin{align}
		Z_n(\bm{\phi}(t)) = R_n(t)\, e^{\mathrm{i}\psi_n(t)} = \frac{1}{N}\sum_{j=1}^N e^{\mathrm{i}n\phi_j(t)}.\label{eq:kuramoto_order}
	\end{align}
The first order parameter $Z_1$ can be thought of as the centroid of the $N$ phases of the oscillators represented on the unit circle, i.e., in the complex plane~\cite{ROD16}. Here, $\psi_n(t)$ is the collective mean phase of the population, and the modulus $R_n(t)$ is given by the absolute value $R_n(t) = \lvert(1/N)\sum_{j=1}^N e^{\mathrm{i}n\phi_j(t)}\rvert$. The quantity $n\in \mathbb{N}$ is also referred to as $n$th moment of the order parameter. Note that here $\mathrm{i} = \sqrt{-1}$ denotes the imaginary unit. In the case of in-phase synchronization, i.e., $\bm{\phi} = (a,\dots,a)^T$ for some $a \in[0,2\pi]$, $R_n = 1$. If $R_n = 0$, we consider the oscillators as incoherent with respect to the $n$th moment.
				
\section{Multicluster states in adaptive networks of phase oscillators}\label{sec:multicluster}
In this section, we show the emergence of different synchronization patterns for an all-to-all coupled base topology $a_{ij}=1$ ($i\ne j$) starting from uniformly distributed random initial conditions ($\phi_i \in [0,2\pi)$, $\kappa_{ij} \in [-1,1]$ for all $i,j = 1,\dots,N$).
	
System~\eqref{eq:adapt_phase_osci1}--\eqref{eq:adapt_phase_osci2} generalizes Kuramoto-Sakaguchi type systems with fixed $\kappa_{ij}$, and has attracted a lot of attention recently \cite{SEL02,MAI07,REN07,AOK09,AOK11,AOK12,PIC11a,TIM14,GUS15a,HA16a,KAS16a,NEK16,KAS17,KAS18,AVA18,KAC20}. In this section, we briefly summarize the findings that have been reported in~\cite{KAS17,BER19,BER19a} and show different types of multicluster states that emerge starting from random initial conditions. These states are characterized by strongly coupled oscillators within each cluster but weak couplings between them. Further, all oscillators in one cluster share a common frequency, while the frequencies between the clusters differ. 
 		
In a multicluster state, the coupling weight matrix with elements $\kappa_{ij}$ can be divided into $M\in\mathbb{N}$ blocks, called clusters, each containing a number $N_\mu$ ($\mu=1,\dots,M$) of frequency synchronized oscillators. We denote the entries of this coupling weight matrix by $\kappa_{ij,\mu\nu}$, referring to the coupling weight from the $j$th oscillator in the $\nu$th cluster to the $i$th oscillator in the $\mu$th cluster. For the temporal behavior of an oscillator in an $M$-cluster state we assume the form
	\begin{align}\label{eq:mc_phases}
		\phi_{i,\mu}(t) = \Omega_\mu t + a_{i,\mu} + s_{i,\mu}(t),
	\end{align}
	where $\phi_{i,\mu}$ denotes the phase of oscillator $i$ inside cluster $\mu$, $\Omega_\mu$ is the collective frequency of the cluster, $a_{i,\mu}\in[0,2\pi)$ are phase lags, and function $s_{i,\mu}(t)$ is considered to be a bounded function due to the interaction of the clusters. 
	
\begin{figure}
		\centering
		\def\svgwidth{0.7\columnwidth}
		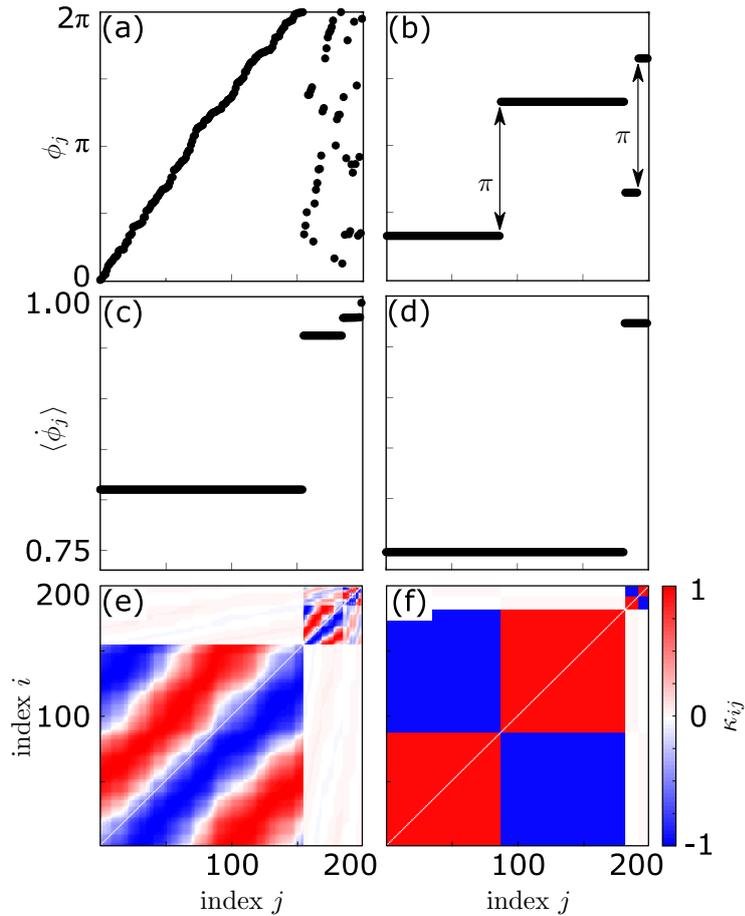
		\caption{Multicluster states in a network of $N=200$ adaptively coupled phase oscillators Eqs.~\eqref{eq:adapt_phase_osci1},\eqref{eq:adapt_phase_osci2}. (a,b): snapshots of the phases $\phi_j$ at $t=20000$; (c,d): mean frequencies $\langle\dot{\phi_j}\rangle = (\phi(t_0+T)-\phi(t_0))/T$ with $t_0 = 10\,000$, $T=10\,000$; (e,f): snapshots of the coupling matrices $\kappa_{ij}$ at $t = 20000$. In (a,c,e) a splay-type multicluster for $\alpha = 0.2\pi,\, \beta = 0.15\pi$ and in (b,d,f) an antipodal-type multicluster with $\alpha = 0.2\pi,\, \beta = -0.6\pi$ are displayed. Further parameters: $\epsilon = 0.01,\, \omega=1,\, \sigma=1/N$.}
		\label{fig:phase_multicluster}
	\end{figure}

	Despite the constant frequencies within a cluster, we can differentiate between three types of multiclusters, depending on the oscillator phases~\cite{BER19,BER19a}. The first type is called \textit{splay-type} multicluster (Fig. \ref{fig:phase_multicluster}(a,c,e)). In this case, the coupling weights $\kappa_{ij,\mu\nu}$ are either constant or change periodically in time, depending on whether the oscillators $\phi_{i,\mu}$ and $\phi_{j,\nu}$ belong to the same ($\mu=\nu$), or a different ($\mu\ne\nu$) cluster, respectively. The amplitude of the coupling weights depends on the frequency difference between the clusters, where a higher frequency difference leads to a smaller amplitude. The separation into three strongly coupled clusters can be seen in Fig.~\ref{fig:phase_multicluster}(a,c,e), as well as the hierarchical structure in the cluster size. Regarding the phases of the oscillators, the oscillators are distributed on the unit circle such that the phases of each splay-type cluster fulfill the condition $R_2(\bm{a}_\mu) = 0$ for $\mu = 1,\,2,\,3$ (cf. Eq.~\eqref{eq:kuramoto_order}). 
	
	Figure~\ref{fig:phase_multicluster}(b,d,f) shows another possible type of multicluster. Here, the oscillators of each cluster possess the phase $a_{i,\mu} = a_{\mu}$ or the antipodal phase $a_{i,\mu} = a_\mu +\pi$, such that $2a_{i,\mu} = 2a_\mu$ for all $i=1,\dots,N_\mu$. Therefore, we call these states \emph{antipodal-type} multicluster. In contrast to splay states, the phase distribution of this type of state fulfills $R_2(\bm{a}_\mu) = 1$, where $\mu = 1,2$. Note that in-phase synchronous states belong to this type of clusters.
	
	 A third possible type of multiclusters combines the previous two types, where the cluster can be either of splay- or antipodal-type. These states are called \emph{mixed-type} multicluster. For more details we refer the reader to~\cite{BER19,BER19a}.
	
\begin{figure}
		\centering
		\def\svgwidth{\columnwidth}
		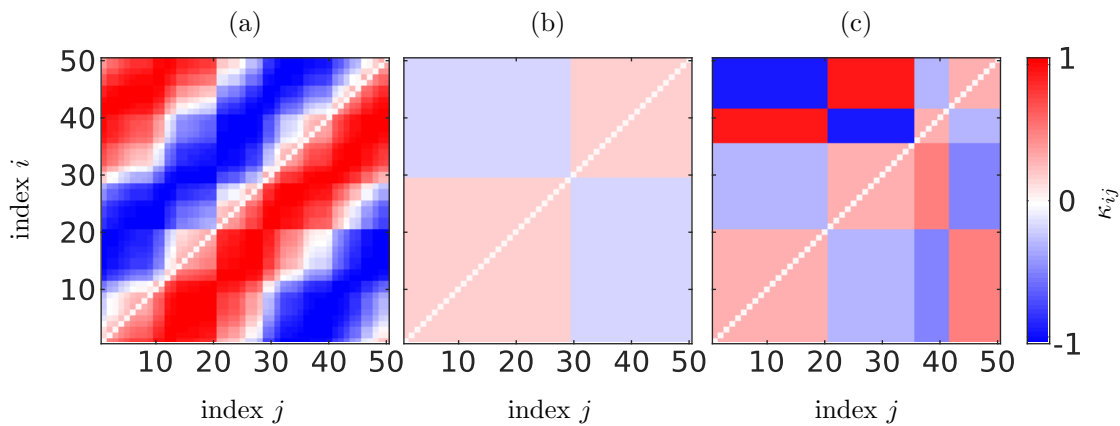
		\caption{Illustration of the coupling weights $\kappa_{ij}$ for all three existing types of one-cluster states for system \eqref{eq:adapt_phase_osci1}--\eqref{eq:adapt_phase_osci2}. (a): splay state with $\alpha = 0.3\pi,\, \beta = 0.1\pi$, (b): antipodal state with $\alpha = 0.2\pi,\, \beta = -0.95\pi$, (c): double-antipodal state with $\alpha = 0.3\pi,\, \beta = -0.15\pi$. Further parameters: $N = 50,\, \epsilon = 0.01,\, \sigma=1/N$. After \cite{BER19a}.
	}
		\label{fig:phase_states}
	\end{figure}

	The appearance of multicluster states suggests that certain one-cluster states serve as building blocks for more complex multicluster states. Formally, a one-cluster state is a frequency synchronized group of phase oscillators, described by
	\begin{align*}
		\phi_i = \Omega\, t + a_i,
	\end{align*}
	with collective frequency $\Omega$, relative phase shifts $a_i \in [0,2\pi)$, and $i = 1,\dots,N$. Figure~\ref{fig:phase_states} shows the coupling matrices $\kappa_{ij}$ of all three possible one-cluster states on an all-to-all network of adaptively coupled phase oscillators \eqref{eq:adapt_phase_osci1}--\eqref{eq:adapt_phase_osci2}. The first two types, namely the splay clusters and antipodal clusters, are described above, and serve as building blocks for multicluster states. The third type, Fig. \ref{fig:phase_states}(c), is called double-antipodal state. This type consists of two groups of antipodal phase oscillators with a fixed phase lag between them. In contrast to splay- and antipodal clusters, double-antipodal states are unstable for the whole range of parameters, and are therefore unlikely to be found as building blocks for multicluster states.
	
\section{Desynchronization of multiclusters by random dilution of network connections}\label{sec:dilution}

	In Section \ref{sec:multicluster}, we have described the generic appearance of multicluster states in system~\eqref{eq:adapt_phase_osci1}--\eqref{eq:adapt_phase_osci2} on an all-to-all coupled network. 
	Based on the oscillators' phase relations within the clusters, we distinguish between antipodal and splay-type multiclusters. We have shown that the oscillators can form groups of strongly connected units, where the interaction between the groups is weak compared to the interaction within the groups. This section investigates the robustness of multicluster states against the random dilution of the underlying network topology.
		
	We consider a network of $N = 100$ adaptively coupled phase oscillators, described by Eqs.~\eqref{eq:adapt_phase_osci1}--\eqref{eq:adapt_phase_osci2}. We fix the parameters $\alpha = 0.3$ and $\beta = -0.53$, such that a multicluster state emerges from random initial conditions for an all-to-all coupled network structure.  In order to implement the dilution of links, we randomly and successively delete $Q$ links ($Q\le (N-1)N$) by choosing a set of $Q$ indices $(ij)$ randomly that correspond to existing links in the adjacency matrix, i.e. $a_{ij} = 1$. These links are then removed, i.e., we set $a_{ij} = 0$. The degree of dilution, i.e., the ratio of deleted links, is defined as $q = Q/(N(N-1))$.
	
	We apply two different numerical approaches to study the effects of dilution: (I) The system~\eqref{eq:adapt_phase_osci1}--\eqref{eq:adapt_phase_osci2} is numerically solved for $11\,000$ time units and successively increasing $q$ in each simulation run, where the final multicluster state for $q=0$ is set as the initial condition for all following simulations. We use this approach to investigate the robustness of a known multicluster state against a random dilution of links. (II) We fix a set of $100$ different random initial conditions. For each $q$ we simulate the system dynamics for all $100$ initial conditions and $11\,000$ time units.
	
\begin{figure}
		\centering
		\def\svgwidth{\columnwidth}
		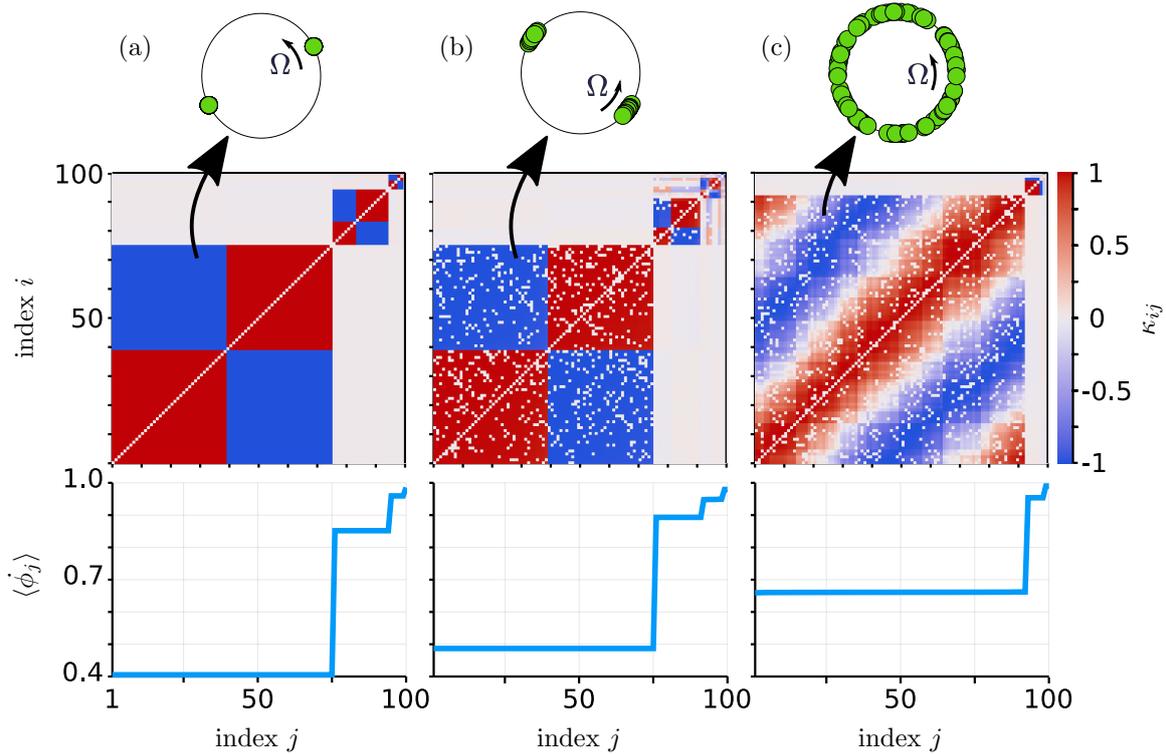
		\caption{Three multicluster states of system \eqref{eq:adapt_phase_osci1}--\eqref{eq:adapt_phase_osci2} for different ratios of deleted links $q$. In the bottom and middle panels, the mean frequencies $\langle\dot{\phi_j}\rangle =(\phi(t_0+T)-\phi(t_0))/T$ with $t_0 = 10\,000$, $T=1000$, and snapshots of the coupling weights $\kappa_{ij}$ at $t = 11\,000$ are shown, respectively. The phases of the oscillators within the biggest cluster at $t = 11\,000$ represented on the unit circle are displayed in the top panel. (a) $q = 0$, (b) and (c) $q = 0.13$ (different initial conditions: (b) multicluster state of panel (a), (c): random initial conditions). Other parameters: $N = 100$, $\epsilon = 0.01$, $\omega = 1$, $\alpha = 0.3\pi$, $\beta = -0.53\pi$, $\sigma=1/N$.}
		\label{fig:phase_coupling}
	\end{figure}

	Figure~\ref{fig:phase_coupling} depicts three resulting states of system~\eqref{eq:adapt_phase_osci1}--\eqref{eq:adapt_phase_osci2}, obtained by the numerical simulations described above. The coupling weights $\kappa_{ij}$ are shown together with the corresponding mean frequencies $\langle\dot{\phi_j}\rangle = (\phi_i(t_0+T)-\phi_i(t_0))/T$, for different values of $q$. Here, we choose $t_0 = 10\,000$ and $T = 1000$. Additionally, the phases $\phi_j$ of the oscillators within the biggest cluster are represented on the unit circle. In the case of a fully coupled network ($q=0$, Fig.~\ref{fig:phase_coupling}(a)), an antipodal-type multicluster state emerges, consisting of three groups. The average frequencies of the oscillators within each cluster are constant, as described in Sec.~\ref{sec:multicluster}. Moreover, the oscillator phases within each cluster possess the phase difference of either $0$ or $\pi$; hence the clusters are of the antipodal type. The smaller clusters possess a mean frequency closer to the natural frequency $\omega=1$ due to their smaller size. 
	
	Figures~\ref{fig:phase_coupling}(b) and \ref{fig:phase_coupling}(c) depict two possible states on networks with dilution degree (ratio of deleted links) $q=0.13$, where the initial conditions were chosen as the multicluster state of Fig.~\ref{fig:phase_coupling}(a) in panel (b) and as random initial conditions in panel (c). Note that the white dots in the coupling matrix represent the deleted links. Our simulations show that despite the missing links, the oscillators can still organize themselves in multiclusters.  The antipodal-type multicluster in Fig.~\ref{fig:phase_coupling}(b) has a similar shape as the multiclusters previously observed. In the case of missing links however, the phases $\phi_i$ within the clusters are slightly spread out in order to compensate for the heterogeneity in the network topology. The splay-type cluster in Fig.~\ref{fig:phase_coupling}(c) is a different type of stable state that can emerge in system \eqref{eq:adapt_phase_osci1}--\eqref{eq:adapt_phase_osci2} for $q=0.13$. Note that in (b) and (c), the same parameter values are used. This shows that different types of multicluster patterns may emerge from random initial conditions. Therefore, we observe that multiclusters may exhibit multistability.
	
\begin{figure}
		\centering
		\def\svgwidth{0.9\columnwidth}
		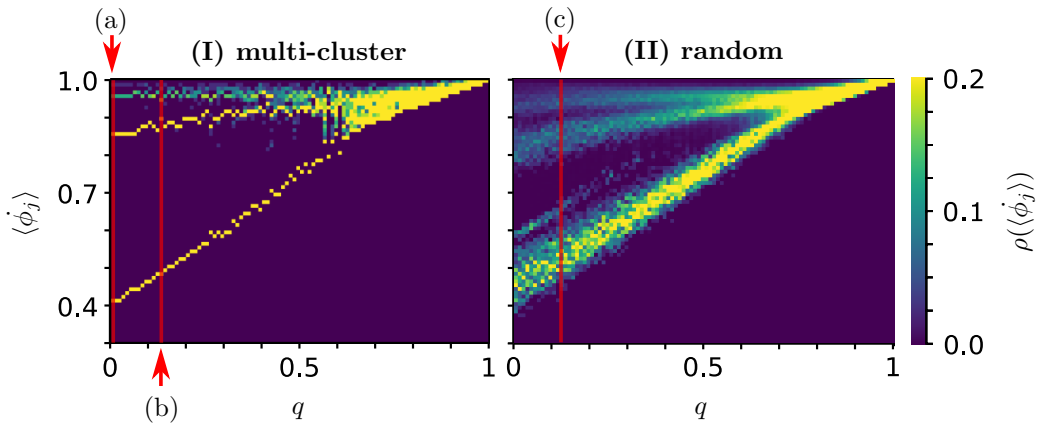
		\caption{Collective mean frequencies $\langle\dot{\phi_j}\rangle$ vs the ratio of deleted links $q$ for system \eqref{eq:adapt_phase_osci1}--\eqref{eq:adapt_phase_osci2}. (I): distribution of mean frequencies  $\langle\dot{\phi_j}\rangle$, where the multicluster state obtained at $q=0$ serves as initial condition for all simulations with $q>0$. (II): distribution of mean frequencies averaged over $100$ realizations of random initial conditions for each $q$. The red lines (a)--(c) mark the states shown in Fig.~\ref{fig:phase_coupling}. Note $\langle\dot{\phi_j}\rangle = (\phi_j(t_0+T)-\phi_j(t_0))/T$, where $t_0 = 10\,000$ and $T = 1000$. Parameters: $N = 100$, $\epsilon = 0.01$, $\omega = 1$, $\alpha = 0.3\pi$, $\beta = -0.53\pi$, $\sigma=1/N$.}
		\label{fig:phase_avrg_freq_statistic}
	\end{figure}

	In order to further investigate the robustness of multiclusters against dilution of the connectivities, we present the distribution $\rho$ (color coded) of the mean frequencies $\langle\dot{\phi_j}\rangle$ of the oscillators versus the fraction of deleted links $q$ in Fig.~\ref{fig:phase_avrg_freq_statistic}. We have obtained these results by applying the two numerical approaches described above. In Fig.~\ref{fig:phase_avrg_freq_statistic}(I) we show the mean frequencies corresponding to the numerical procedure (I) where the multicluster state depicted in Fig. \ref{fig:phase_coupling}(a) is set as initial condition. Note that here the data for each step of $q$ is an average of all $N$ oscillators. For a wide range of $q$, three distinct clusters are visible. For large values of $q$, where the network becomes sparse, the frequency clusters are not clearly separated. The qualitative shape of the initial multicluster state is preserved on networks of different degrees of dilution, ranging from fully coupled to almost uncoupled topologies. The number of clusters stays the same on networks with varying numbers of links, however, the collective mean frequencies of the oscillators adapt to the changes in the coupling topology, in a linear relation with $q$.
	
	The linear dependence of the cluster frequencies on the relative number of deleted links $q$ can be explained as follows. For increasing dilution $q$, each link has an equal probability to be cut off. Therefore, on average, in each cluster there exist $(1-q) \cdot N_\mu(N_\mu-1)$ links. Furthermore, by assuming $\epsilon\ll 1$, the collective frequency of each cluster can be roughly approximated up to zeroth order in $\epsilon$ by $\Omega_\mu\approx \omega + \sigma\sum_{j=1}^{N_\mu}a_{ij,\mu\mu}\sin(a_{i,\mu}-a_{j,\mu}+\beta)\sin(a_{i,\mu}-a_{j,\mu}+\alpha) $~\cite{BER19}. Let us consider an approximately antipodal cluster, i.e., $a_{i,\mu}-a_{j,\mu}\approx 0$ or $\pi$. Then $\Omega_\mu\approx \omega + \sigma\sin(\beta)\sin(\alpha)r_{i,\mu}$ where $r_{i,\mu}=\sum_{j=1}^{N_\mu}a_{ij,\mu\mu}$ is the $i$th row sum restricted to the $\mu$th cluster. By averaging over $i=1,\dots,N_\mu$, we end up with the approximation 
	\begin{align}\label{eq:clusterFreqApprox}
		\Omega_\mu\approx \omega + \sigma (1-q) (N_\mu-1) \sin(\beta)\sin(\alpha).
	\end{align}
The latter expression explains the linear dependence of the cluster frequency on the ratio of deleted links. Furthermore, Eq.~\eqref{eq:clusterFreqApprox} shows that the slope of the linear relation depends on the cluster size. This is in agreement with the findings in Fig.~\ref{fig:phase_avrg_freq_statistic}(I). Note that splay clusters can be treated similarly.
	
	In Fig.~\ref{fig:phase_avrg_freq_statistic}(II), we show the distribution of the collective mean frequencies versus $q$ for random initial conditions according to the numerical procedure (II). Note that here the data for each step of $q$ is an average of $100$ numerical runs. Hence the color code represents the fraction of oscillators that lie in a corresponding frequency band. For a wide range of $q$, the frequencies roughly show three maxima of $\rho(\langle\dot{\phi_j}\rangle)$, indicating 3-cluster states, see also Fig.~\ref{fig:phase_coupling_app} in the Appendix as an example. Further, we observe 2-cluster states of splay type, corresponding to a fourth, smaller maximum of $\rho(\langle\dot{\phi_j}\rangle)$, which vanishes for $q>0.4$ and is located slightly above the largest maximum. For increasing values of $q$, the overall frequencies increase linearly, and eventually converge to $\left.\langle\dot{\phi_j}\rangle\right|_{q=1} = 1$. In this case, all the nodes are uncoupled, and oscillate with their natural frequency of $\Omega = 1$. 
	These results can be compared with reference~\cite{KAS18a}, where the dynamical states on networks of adaptively coupled phase oscillators are studied in the $(\alpha,\beta)$ parameter space for different degree of network dilution. There it has been shown that splay- and antipodal-type clusters emerge in the corresponding region of the parameter space for all-to-all coupled networks. For decreasing number of links, i.e., increasing $q$, the network first loses its ability to reach splay-type states, and subsequently loses its ability to synchronize at all. These findings suggest that the fourth maximum in Fig.~\ref{fig:phase_avrg_freq_statistic}(II) exists due to the existence of splay-type multiclusters since it vanishes for $q>0.4$.
	
\begin{figure}
		\centering
		\def\svgwidth{\columnwidth}
		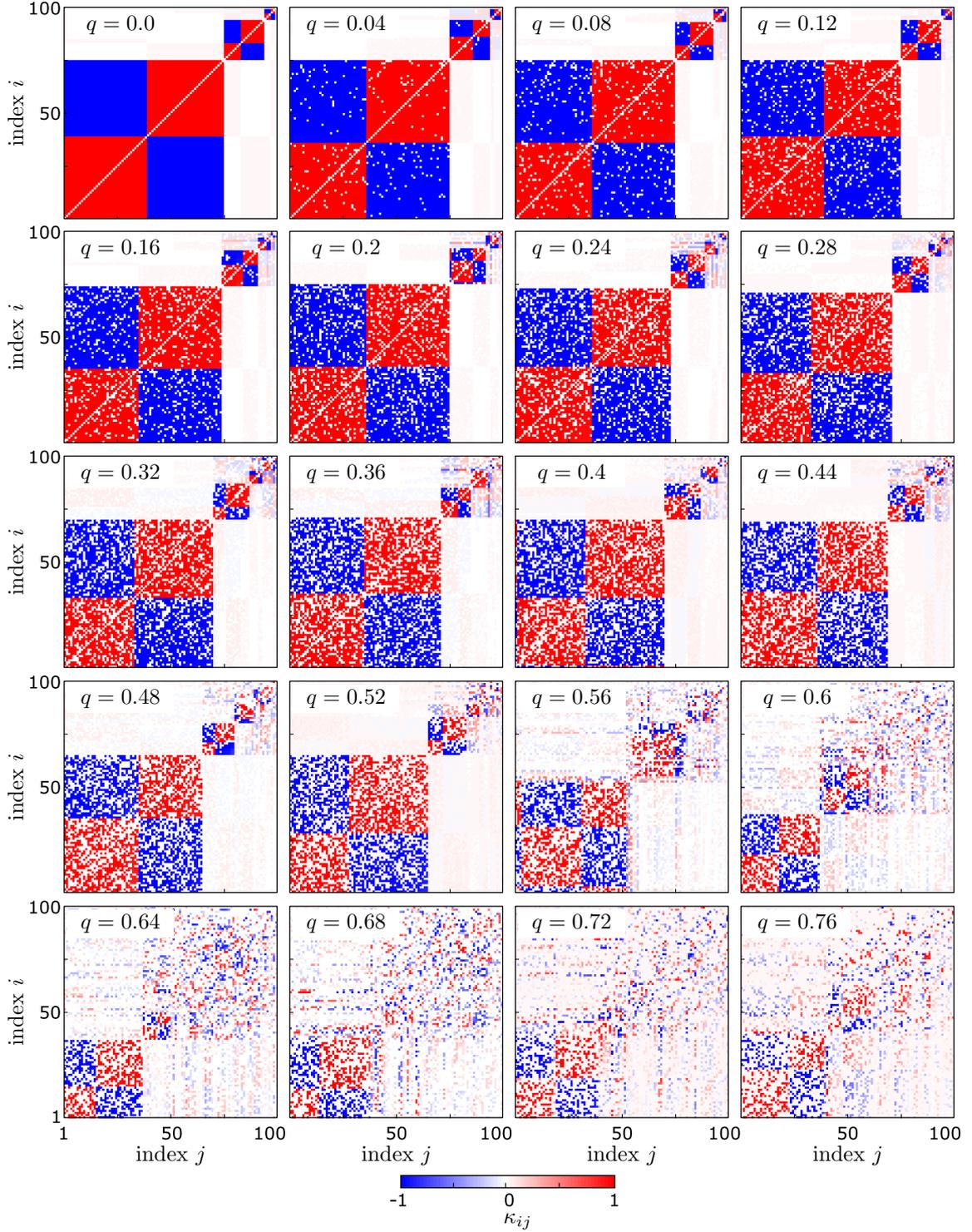
		\caption{Multicluster synchronization in system \eqref{eq:adapt_phase_osci1}--\eqref{eq:adapt_phase_osci2} with decreasing ratio of randomly deleted links $q=Q/(N(N-1))$. The multicluster state at $q=0$ is used as initial condition for all simulations with $q>0$. Snapshots of the coupling matrices $\kappa_{ij}$ at $t=11\,000$ for different values of $q$ are presented. In each panel the oscillators are ordered according to their mean frequency, and subsequently their phases. Parameters: $\alpha = 0.3\pi$, $\beta = -0.53\pi$, $\omega = 1$, $\epsilon = 0.01$, $\sigma=1/N$, $N=100$.}
		\label{fig:phase_coupling_cont}
	\end{figure}

	We visualize the multicluster states described in Fig.~\ref{fig:phase_avrg_freq_statistic}(I) by showing the corresponding snapshots of the coupling matrices $\kappa_{ij}$ in Fig.~\ref{fig:phase_coupling_cont}. We note that the multicluster state at $q=0$ was chosen as the initial condition for all simulations with $q>0$. In agreement with our prior observations, three clusters are visible for densely connected networks. For increasing $q$, meaning sparser networks, the frequency clusters dissolve in a hierarchical manner, where clusters consisting of fewer oscillators vanish prior to those containing a large number of oscillators. Further, we observe that the dissolution of a cluster starts with the uncoupling of single oscillators from the cluster and a continuous decrease in cluster size. This process continues until the cluster vanishes. A similar plot corresponding to Fig.~\ref{fig:phase_avrg_freq_statistic}(II) with random initial conditions can be found in the Appendix, see Fig.~\ref{fig:phase_coupling_app}.
	
	In this section, we have shown that randomly diluting the network topology leads to a desynchronization of multiclusters. It has been observed that system~\eqref{eq:adapt_phase_osci1}--\eqref{eq:adapt_phase_osci2} preserves the qualitative shape of multicluster states on random networks of different sparsity. Hence, multiclusters are robust against topological perturbations. Further, system~\eqref{eq:adapt_phase_osci1}--\eqref{eq:adapt_phase_osci2} possesses a high degree of multistability also for diluted network topologies. In the following section, we investigate the stability of synchronized states by means of the master stability function for networks with adaptive couplings~\cite{BER20b}. With this we show that the presence of adaptive couplings can be used to influence the stability of system~\eqref{eq:adapt_phase_osci1}--\eqref{eq:adapt_phase_osci2}. Further, we show that the destabilization of the phase-synchronized state has implications for the emergence of multicluster states.

\section{Master stability function for adaptive phase oscillator networks}\label{sec:msf_phase_osci}

In the preceding section, we have described the desynchronization of multicluster states for a decreasing number of links. It has been shown that system~\eqref{eq:adapt_phase_osci1}--\eqref{eq:adapt_phase_osci2} keeps its ability to form strongly coupled groups of oscillators from random initial conditions on networks with increasing degree of dilution, while the fraction of desynchronized oscillators increases for sparse networks.  Since the individual clusters in a multicluster state are effectively uncoupled from each other, the stability of these strongly coupled subnetworks may play an important role in the stability of multicluster states. 

In this and the next section, we aim to gain analytic insights into the stability of in-phase synchronized states, i.e., $\phi_i=\Omega t$, $i=1,\dots,N$. Note that the in-phase synchronized states belong to the class of antipodal states, see Sec. \ref{sec:multicluster}, and share the same dynamical properties~\cite{BER19,BER19a}. We assume that the network topology expressed by the adjacency matrix possesses a constant row sum $r$, i.e., $r=\sum_{j = 1}^N a_{ij}$ for all $i$. Following the master stability approach for networks with adaptive coupling weights as developed in~\cite{BER20b}, we derive the master stability function for networks of adaptively coupled phase oscillators. This allows us to separate the effects of the local node dynamics from the effects of adaptivity and from the network topology. Hence, we are able to draw general conclusions on the stability of the in-phase synchronized state, for almost arbitrary complex network topologies. With this, the master stability approach allows us to study the interplay of nodal dynamics, adaptivity, and complex network structures. The approach, further, enables us to control the stability of synchronized states, depending on the changes in the coupling structure.

In the following, we provide a brief discussion of the master stability function for the adaptive Kuramoto-Sakaguchi network \eqref{eq:adapt_phase_osci1}--\eqref{eq:adapt_phase_osci2}. Using the results from~\cite{BER20b}, the stability of the synchronous state of system~\eqref{eq:adapt_phase_osci1}--\eqref{eq:adapt_phase_osci2} is governed by the two linearized differential equations for perturbations of the synchronous state in the new coordinates $\zeta \in\mathbb{R}$ and $\kappa\in\mathbb{R}$ 
\begin{align}\label{eq:MSF_AKSSystem}
\frac{\mathrm{d}}{\mathrm{d}t} \begin{pmatrix}
\zeta \\
\kappa
\end{pmatrix}&= \begin{pmatrix}
\mu\sigma{\cos(\alpha)\sin(\beta)} &  - \sigma{\sin(\alpha)}\\
-\epsilon\mu \cos(\beta) & -\epsilon
\end{pmatrix}\begin{pmatrix}
\zeta \\
\kappa
\end{pmatrix},
\end{align}
where $\mu\in \mathbb{C}$ denotes all eigenvalues of the Laplacian matrix $L=r\mathbb{I}_N -A$ corresponding to the network described by the adjacency matrix $A$. Here, $\mathbb{I}_N$ is the $N$-dimensional identity matrix. The characteristic polynomial in $\lambda$ of~\eqref{eq:MSF_AKSSystem} is of degree two and reads
\begin{align}
	\lambda^2 + \left(\epsilon-\sigma{\mu}\cos(\alpha)\sin(\beta)\right)\lambda -\epsilon\sigma{\mu}\sin(\alpha+\beta) = 0. \label{eq:charpoly}
\end{align}
The master stability function is given by $\Lambda(\sigma\mu)=\max(\mathrm{Re}(\lambda_1),\mathrm{Re}(\lambda_2))$ where $\lambda_1$ and $\lambda_2$ are the two solutions of the quadratic polynomial \eqref{eq:charpoly}. Using the master stability function, we can deduce the local stability of the in-phase state (also for the antipodal states) directly from the connectivity structure given by the adjacency matrix. In particular, if there exists at least one Laplacian eigenvalue such that $\Lambda(\sigma\mu)>0$, the state is locally unstable. If for all Laplacian eigenvalues $\Lambda(\sigma\mu)<0$ holds, the in-phase synchronous state is locally stable. Note that by construction of the Laplacian matrix $L$ there exists always one eigenvalue $\mu=0$ which leads to $\lambda_1=0$ and $\lambda_2=-\epsilon$. The zero eigenvalue of the matrix in~\eqref{eq:MSF_AKSSystem} corresponds to the shift symmetry of system~\eqref{eq:adapt_phase_osci1}--\eqref{eq:adapt_phase_osci2} and is not considered in the above stability condition.

In order to get insight into the form of the master stability function, we consider the boundary of the region in $\sigma{\mu}$ parameter space that corresponds to stable local dynamics. The boundary is given by $\lambda=\mathrm{i}\gamma$ with $\gamma \in \mathbb{R}$. Plugging this into equation~\eqref{eq:charpoly}, we obtain the function
\begin{align*}
\sigma\mu =a(\gamma)+\mathrm{i}b(\gamma) 
\end{align*}
with
\begin{align*}
a(\gamma) &= \epsilon\frac{\gamma^2\left(\cos\alpha\sin\beta-\sin(\alpha+\beta)\right)}{{\gamma}^2\cos^2\alpha\sin^2\beta+\epsilon^2\sin^2(\alpha+\beta)}, \\
b(\gamma) &=\frac{\gamma^3\cos\alpha\sin\beta+\epsilon^2\gamma\sin(\alpha+\beta)}{{\gamma}^2\cos^2\alpha\sin^2\beta+\epsilon^2\sin^2(\alpha+\beta)}. 
\end{align*}
Hence, the boundary in the complex $\sigma\mu$-plane is parametrically described as a cubic function of the real value $\gamma$. Due to the symmetry of the master stability function, a condition to observe a nontrivial shape of the boundary is that the function $\sigma\mu({\gamma})$ possesses three crossings with the real axis, i.e., two positive real solutions for $b(\gamma)=0$. The three crossings are given by $\gamma_1=0$ and as real solutions $\gamma_2$ and $\gamma_3$ of $\gamma^2\cos\alpha\sin\beta=-\epsilon^2\sin(\alpha+\beta)$. From this, we deduce the existence condition for three crossings as $\sin(\alpha+\beta)/(\cos\alpha\sin\beta)<0$ ($\epsilon>0$). Note that $a(\gamma_2)=a(\gamma_3)$. If there are three crossings, the master stability function possesses a stability island in the complex plane. A phenomenon induced by the existence of a stability island is in the focus of the next section. 

\section{Destabilization of synchrony by changes in network connectivity}\label{sec:desync_top_change}

The emergence of stability islands in the master stability function enables us to destabilize synchronous states in various ways. In the following, we demonstrate a desynchronization by modifying the network topology. A similar approach has been studied on non-adaptive, delay-coupled systems in~\cite{LEH11,KEA12}. There it is shown that randomly added inhibitory links lead to a desynchronization transition on regular ring topologies. We extend these studies towards adaptively coupled systems, and use random network topologies.

We consider a network of adaptively coupled phase oscillators, described by the set of differential equations~\eqref{eq:adapt_phase_osci1}--\eqref{eq:adapt_phase_osci2}. The oscillators are assumed to have a natural frequency $\omega = 1$. In the following, we investigate the stability of phase-synchronized states on networks with random adjacency matrices, first by using the master stability function, and second by numerical integration. We choose the parameters $\alpha$ and $\beta$, such that there exists a stability island of the master stability function in the complex plane. We then prepare random initial conditions, i.e., $\phi_i(0) \in [0,2\pi)$ and $\kappa_{ij}(0)\in [-1,1]$, and numerically integrate equations~\eqref{eq:adapt_phase_osci1}--\eqref{eq:adapt_phase_osci2} with $N=100$ and $t = 30000$ for three different random adjacency matrices, corresponding to directed, connected networks with different node in-degree. In order to guarantee the existence of synchronized states, we assume that the adjacency matrices have a constant row sum $r = \sum_{j = 1}^N a_{ij}$. Note that this condition is not preserved by the dilution procedure used in Sect.~\ref{sec:multicluster}. We apply the following procedure to construct the directed random networks. For each node $i$ of the $N$ nodes, $r$ links are randomly picked from the set that consists of all possible links from nodes $j \neq i$ to node $i$. For the selected links $a_{ij}$ are set to $1$. This procedure results in a directed random network with $N$ nodes and a constant row sum (in-degree) $r$. We note that the row sum $r$ defines the ratio of deleted links $q=Q/(N(N-1))=1-r/(N-1)$, i.e., the degree of dilution.

\begin{figure}
	\centering
	\def\svgwidth{0.9\columnwidth}
	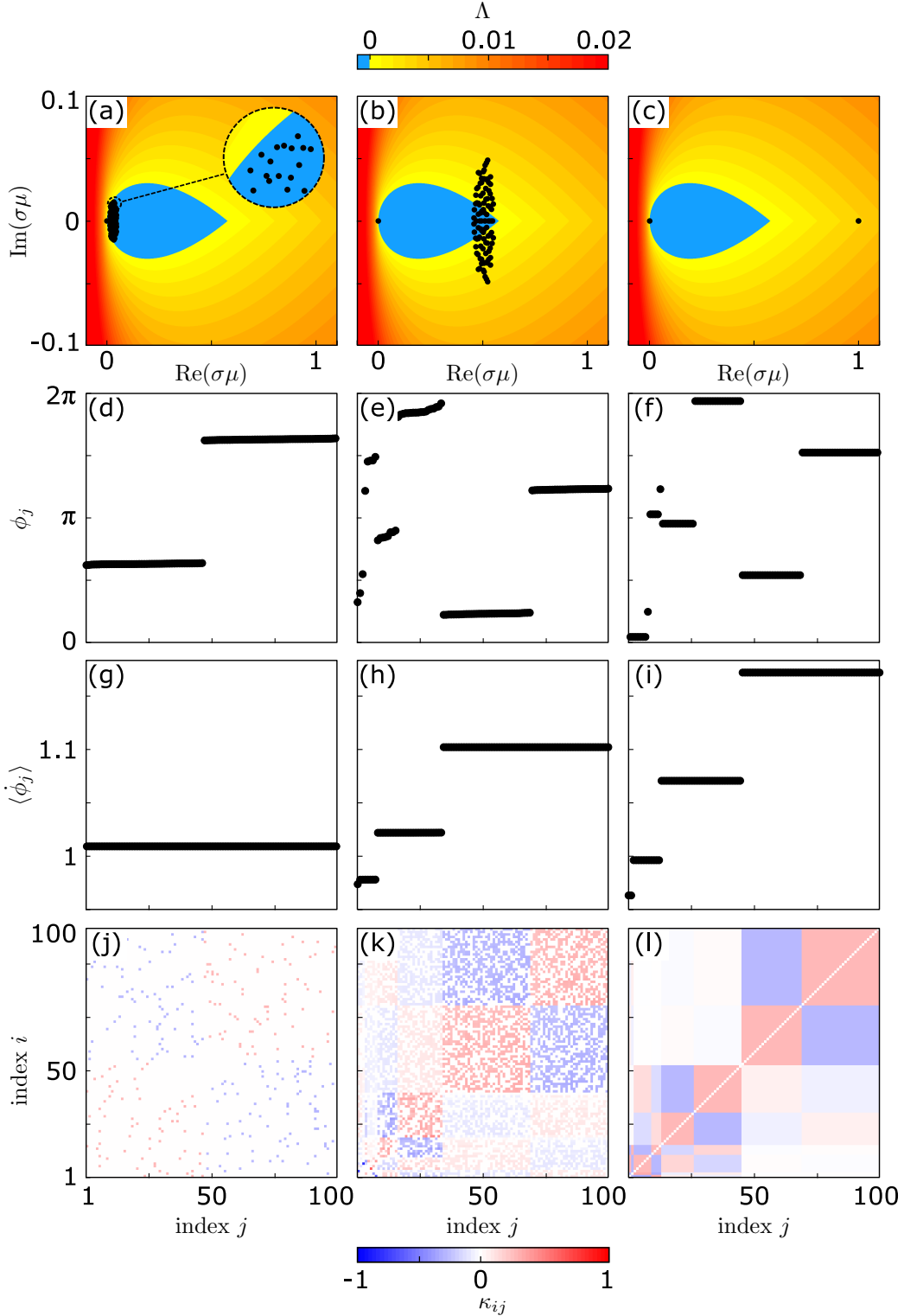
	\caption{Dynamics of a network of $N=100$ adaptively coupled Kuramoto-Sakaguchi oscillators with random adjacency matrices with different constant row sums $r = \sum_{j=1}^N a_{ij}$, and random initial conditions $\phi_i(0) \in [0,2\pi\}$ and $\kappa_{ij}(0) \in [-1,1]$ for $i,j = 1\dots N$. The simulations results are shown for the three values (a,d,g,j) $r = 3$, (b,e,h,k) $r=50$, (c,f,i,l) $r = 99$. The panels show: (a,b,c) the master stability function color coded, together with $\sigma\mu_i$, where $\mu_i$ are the $N$ Laplacian eigenvalues corresponding to each adjacency matrix; the inset in (a) depicts a blow-up of the marked area, where eigenvalues lie close to the border of the stability island; in (d,e,f) snapshots of the oscillators phases $\phi_j$ at $t = 30000$; in (g,h,i) the mean frequencies $\langle\dot{\phi_j}\rangle = (\phi_j(t_0+T)-\phi_j(t_0))/T$, where $t_0 = 25\,000$ and $T = 5000$; in (j,k,l) snapshots of the coupling weights $\kappa_{ij}$, where the coupling weights are color coded. The oscillators are ordered according to their mean frequencies $\langle\dot{\phi_j}\rangle$, and subsequently their phases $\phi_j$. Other parameters: $\alpha = 0.48\pi$, $\beta =  0.91\pi$, $\omega = 1$, $\sigma = 1/N$, $\epsilon = 0.01$.}
	\label{fig:msf_desync}
\end{figure}

In Fig.~\ref{fig:msf_desync}(a,b,c) we show the master stability function $\Lambda(\sigma\mu)$ for the parameters $\alpha = 0.48\pi$ and $\beta = 0.91\pi$, where the black dots indicate the scaled Laplacian eigenvalues $\sigma\mu_i$ of three random adjacency matrices with different row sums $r$. We use this to determine the stability of the in-phase synchronized state for the given coupling topology, and show the corresponding numerical results below.

We observe that for sparse random networks with  $r = 3$ ($q = 0.97$), the synchronous state is stable, see Fig. \ref{fig:msf_desync}(a,d,g,j). The stability follows directly from the master stability function, since all values $\sigma\mu_i$ for all Laplacian eigenvalues lie within the stable region $\Lambda(\sigma\mu)<0$ (blue), see Fig. \ref{fig:msf_desync}(a). In this case, all oscillators have the same mean frequency $\langle\dot{\phi_j}\rangle$, see Fig. \ref{fig:msf_desync}(g). Moreover, the oscillators $\phi_j$ either share the same phase $a_i \approx a$, or the antipodal phase $a_i \approx a+\pi$, indicating an antipodal-type cluster. We note that the in-phase synchronous states considered by the master stability function $\Lambda$ belong to the class of antipodal states and share the same local stability properties~\cite{BER19a}. If the number of links is increased, e.g. to $r = 50$ ($q = 0.49$), some of the Laplacian eigenvalues move out of the stable region of the master stability function into the unstable region (yellow/red) in the complex plane. Consequently the in-phase synchronized state is locally unstable, see Fig.~\ref{fig:msf_desync}(b). In this case, we observe the emergence of antipodal-type multiclusters, where the oscillators within each cluster share a common frequency, Fig. \ref{fig:msf_desync}(h), and have antipodal phase relations, Fig.~\ref{fig:msf_desync}(e). We note that the phases within the two smallest clusters are spread out, due to the fact that the adjacency matrix restricted to the clusters does not necessarily have a constant row sum. In the case of an all-to-all coupled network with $r = 99$ ($q = 0$), the values $\sigma\mu_i$ are either located at $0$ or $1$, see Fig.~\ref{fig:msf_desync}(c). Since the eigenvalues located at $\sigma\mu = 1$ lie in the unstable region, the in-phase synchronized state is unstable. In this case, we again observe the emergence of an antipodal-type multicluster, see Fig. \ref{fig:msf_desync}(f,i,l).

In this section, we have presented the effects induced by changes of the network topology on the desynchronization of the phase-synchronized state. While the phase-synchronized state is stable for sparse networks, it is destabilized for an increasing number of links. We have described this behavior by means of the master stability function for networks with adaptive coupling weights. Modifying the adjacency matrix affects the Laplacian eigenvalues $\mu_i$, and consequently changes the largest Lyapunov exponents of the network. We use the presence of bounded stable regions (stability islands in the complex plane) of the master stability function in order to successively shift the Laplacian eigenvalues from the stable into the unstable regime by changing the network connectivities. We show that the resulting desynchronization has strong implications for the emergence of multicluster states.

\section{Conclusion}

We have investigated the emergence of cluster synchronization on networks of adaptively coupled Kuramoto-Sakaguchi oscillators with complex topologies. Specifically, we have focused on the robustness of the multifrequency cluster states against diluted connectivities. We have shown the influence of topological changes on these states. Further, we have complemented these studies by an analytical approach describing the stability of phase-synchronized states. Investigating the master stability function for adaptive Kuramoto-Sakaguchi networks has allowed us to study the interplay between nodal dynamics, adaptivity, and network topology. Using this, we have shown that the existence of adaptive coupling weights may dramatically change the synchronization behavior with regard to network topology.

In Sec.~\ref{sec:multicluster}, we have reviewed previous findings on the emergence of multi-frequency-clusters on networks of adaptively coupled phase oscillators, and introduced a notation describing the generic appearance of cluster states. We have illustrated numerically for all-to-all coupled networks that, starting from random initial conditions, these systems can reach different types of multicluster states, such as splay- and antipodal-type multiclusters. These states have previously been studied on all-to-all coupled networks~\cite{KAS17,BER19,BER19a}, nonlocally coupled rings~\cite{BER20c}, multiplex~\cite{KAS18,BER20} and random networks~\cite{KAS18a} with adaptive coupling weights. We have extended these studies towards randomly diluted coupling topologies. In order to investigate the influence of topological changes upon these states, we have modified the adjacency matrix describing the underlying time-independent coupling topology. For this purpose we have constructed random adjacency matrices, representing directed random network topologies with varying degree of dilution.

We have shown in Sec. \ref{sec:dilution} that the emergence of partially synchronized states is maintained on random networks also with a small number of links. The emergence of different types of multicluster states including splay-type multiclusters has been observed on densely coupled networks ($q=0.13$ in Fig.~\ref{fig:phase_coupling}(c)). This result is in agreement with the findings presented in~\cite{KAS18a}, where the dynamical states on networks of adaptively coupled phase oscillators have been studied in the $(\alpha,\beta)$ parameter space for different network sparsities. We have shown in numerical investigations that the multicluster states observed are multistable with regard to initial conditions, meaning that different multiclusters may emerge for different initial conditions. By depicting asymptotic states for different values of the ratio of deleted links $q$, we have shown that the qualitative shape of a given multicluster state is preserved on networks of different degree of dilution. The number of clusters stays the same for a wide range of $q$, however, the frequencies of the oscillators adapt to the changes in the coupling topology, in a linear relation with $q$. The latter effect has been analytically described.

Since in-phase synchronous and antipodal states have the same local stability properties, we have extended our investigations by an analytical approach describing the stability of in-phase synchronized states in Sec.~\ref{sec:msf_phase_osci}. We have analyzed the master stability function for networks of Kuramoto-Sakaguchi oscillators with adaptive coupling weights using the novel methods presented in~\cite{BER20b}. We have observed the emergence of bounded regions that lead to stable synchronous dynamics in the master stability function, representing stability islands in the complex plane. We have analytically described the stability border, and provided a condition for the emergence of stability islands. Due to the shape of these stability islands, it is possible to destabilize in-phase synchronized states by increasing the number of links within the network. We have shown that such a destabilization has implications for the emergence of multicluster states. By tailoring the system configurations such that a stability island emerges in the master stability function, we have observed stable multicluster states emerging from random initial conditions for those topological configuration with unstable in-phase synchronized states. In previous work~\cite{BER20b}, it has been shown that such a counterintuitive desynchronization effect is also possible by increasing the overall coupling strength.

In this work, we have shown the emergence of multicluster states on networks of adaptively coupled Kuramoto-Sakaguchi oscillators with random coupling topologies. Due to the adaptivity of the coupling weights, the individual oscillators are able to adapt their frequencies and form strongly coupled groups of frequency-synchronized units. We have shown by numerical and analytical investigations how random coupling topologies affect the emergence of these synchronization patterns. While the structural shape of multiclusters stays the same, the overall synchrony in a network declines with decreasing number of links, since more and more oscillators decouple from the system and exhibit incoherent dynamics. Nonetheless, we have shown by means of the master stability function that some network configurations allow for stable in-phase synchronized dynamics for very sparse random networks. While our investigations are restricted to networks with uniform degree distributions, they may be generalized to more realistic network models, such as small-world or scale-free networks. The master stability approach serves as a universal tool, describing the stability of in-phase synchronized states. We have shown that the destabilization of fully synchronized states has implications for the emergence of multicluster states. In order to obtain a more complete picture, this approach may be extended towards splay-type synchrony. Our findings on the effects of diluted connectivities upon multicluster states illustrate the complex interplay between topology and adaptivity. By applying the master stability approach, we have shown that the adaptive properties of a network can have a huge impact on the stability of fully and partially synchronized states since the presence of adaptivity provides a feedback mechanism that can change the stability; intuitively this is similar to an additional effective phase lag.

\section*{Acknowledgments}
	This work was supported by the German Research Foundation DFG, Project Nos. 411803875 and 440145547.

\begin{appendices}
	\section{Appendix}\label{sec:appendix}
\begin{figure}
		\centering
		\def\svgwidth{\columnwidth}
		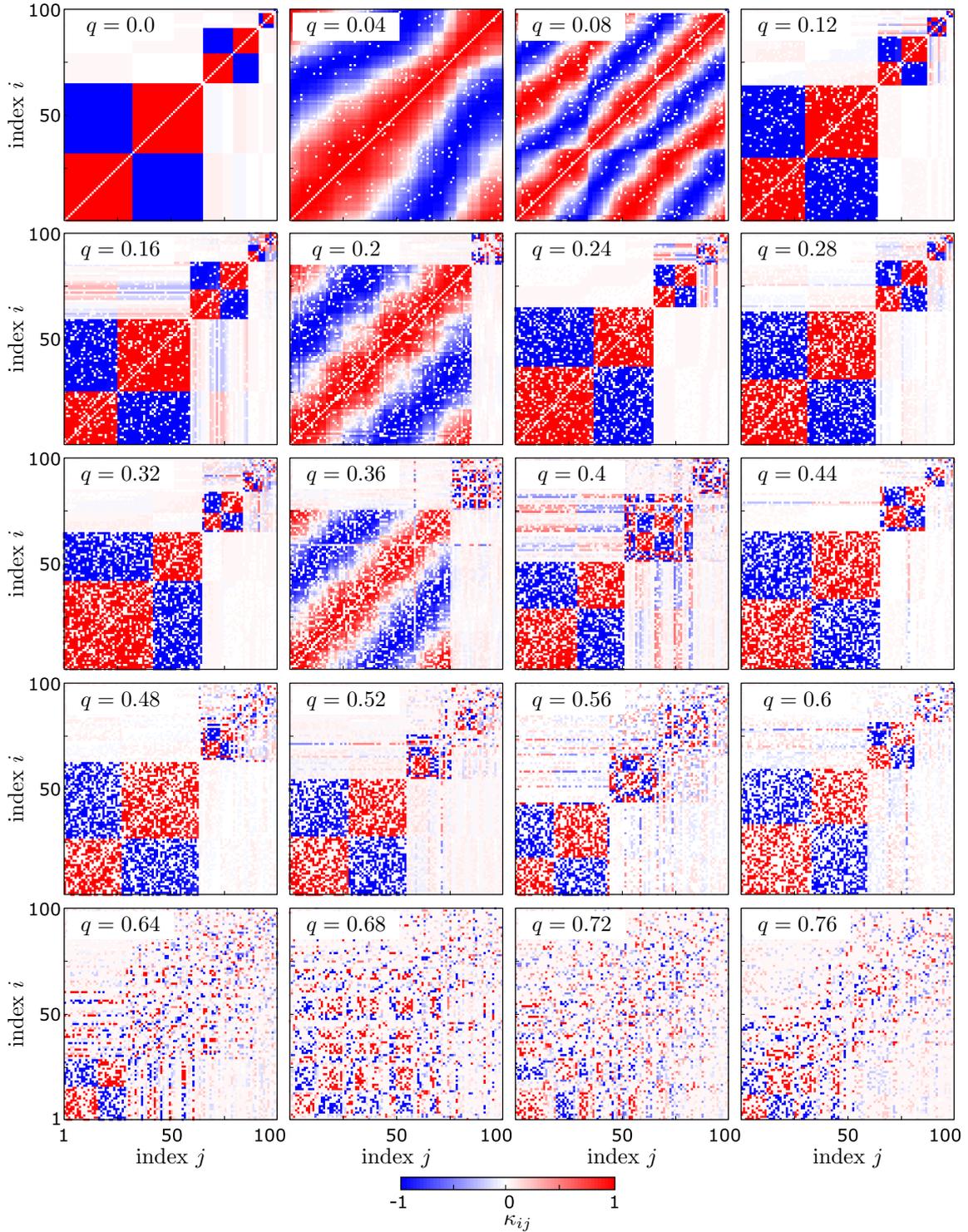
		\caption{Multicluster synchronization on system \eqref{eq:adapt_phase_osci1}--\eqref{eq:adapt_phase_osci2} with increasing ratio of randomly deleted links $q$ and random initial conditions. Snapshots of the coupling matrices $\kappa_{ij}$ at $t=11\,000$ for different values of $q$ are presented. In each panel the oscillators are ordered according to their average frequency, and subsequently the phases. Parameters: $\alpha = 0.3\pi$, $\beta = -0.53\pi$, $\omega = 1$, $\epsilon = 0.01$, $N=100$.}		
		\label{fig:phase_coupling_app}
	\end{figure}

Figure~\ref{fig:phase_coupling}(b,c) shows that system~\eqref{eq:adapt_phase_osci1}--\eqref{eq:adapt_phase_osci2} is multistable with regard to initial conditions. This means that depending upon the initial state different realizations of multicluster states can emerge. We illustrate this feature in Fig. \ref{fig:phase_coupling_app}, where we show the coupling matrices $\kappa_{ij}$ for different values of $q$. Note that the presented coupling matrices correspond to the dynamics shown in Fig.~\ref{fig:phase_avrg_freq_statistic}(II).
\end{appendices}

\bibliography{ref}
\bibliographystyle{prwithtitle}

\end{document}